\documentclass[%
reprint,
showpacs,
bibnotes,
amsmath,amssymb,
aps,
prb,
floatfix,
]{revtex4-1}

\usepackage{graphicx}
\usepackage{dcolumn}
\usepackage{bm}
\usepackage{multirow}
\usepackage{array}
\usepackage{booktabs}
\usepackage{ctable}
\usepackage{upgreek}
\usepackage{epsfig}
\usepackage{mathrsfs}
\usepackage{amssymb}
\usepackage{amsbsy}
\usepackage{color}
\usepackage{cancel}
\usepackage{marginnote}
\newcommand{\bcen}{\begin{center}}
\newcommand{\ecen}{\end{center}}
\newcommand{\btab}{\begin{tabular}}
\newcommand{\etab}{\end{tabular}}
\newcommand{\bdes}{\begin{description}}
\newcommand{\edes}{\end{description}}

\newcommand{\beq}{\begin{equation}}
\newcommand{\eeq}{\end{equation}}
\newcommand{\bea}{\begin{eqnarray}}
\newcommand{\eea}{\end{eqnarray}}

\newcommand{\half}{\frac{1}{2}}
\newcommand{\bary}{\begin{array}}
\newcommand{\eary}{\end{array}}
\newcommand{\benum}{\begin{enumerate}}
\newcommand{\eenum}{\end{enumerate}}
\newcommand{\bitem}{\begin{itemize}}
\newcommand{\eitem}{\end{itemize}}

\newcommand{\btau}{\mbox{\boldmath $ \tau $}}
\newcommand{\blam}{{\boldsymbol{\lambda}}}

\newcommand{\bOne}{{\boldsymbol 1}}
\newcommand{\be} { \mbox{\boldmath $e$}}

\newcommand{\bk} { \bm{k} }

\newcommand{\bp} { \bm{p} }
\newcommand{\bq} { \bm{q} }
\newcommand{\br} { \boldsymbol{r}}

\newcommand{\bA} { \mbox{\boldmath $A$}}

\newcommand{\bzero} { {\boldsymbol{0}}}

\newcommand{\D}[1]{\mbox{d}{#1}}

\newcommand{\mean}[1]{\langle #1 \rangle}

\newcommand{\ket}[1]{| #1 \rangle}
\newcommand{\braket}[2]{\langle #1 | #2 \rangle}

\newcommand{\eqn}[1] {eqn.~(\ref{#1})}
\newcommand{\prn}[1] {(\ref{#1})}

\newcommand{\fig}[1]{fig.~\ref{#1}}
\newcommand{\Fig}[1]{Fig.~\ref{#1}}

\makeatletter

\newcommand{\Rmnum}[1]{\expandafter\@slowromancap\romannumeral #1@}
\makeatother

\newcommand{\myfigwidth}{0.99\columnwidth}

\newcommand{\asc}{a_{sc}}
\newcommand{\as}{a_{s}}
\newcommand{\Eb}{E_{b}}
\newcommand{\Ef}{E_F}
\newcommand{\kf}{k_F}

\newcommand{\etaT}{{\eta_t}}
\newcommand{\muNI}{{\mu_{NI}}}

\newcommand{\mylabel}[1]{\label{#1}} 

\newcommand{\myonlinecite}[1]{[\onlinecite{#1}]}

\usepackage{lineno}

\begin{document}



\title{BCS-BEC crossover induced by a synthetic non-Abelian gauge field}

\author{Jayantha P. Vyasanakere$^1$}
\email{jayantha@physics.iisc.ernet.in}
\author{Shizhong Zhang$^2$}
\email{shizhong.zhang@gmail.com}
\author{Vijay B. Shenoy$^1$}
\email{shenoy@physics.iisc.ernet.in}

\affiliation{$^1$Centre for Condensed Matter Theory, Department of Physics, Indian Institute of Science, Bangalore 560 012, India}

\affiliation{$^2$Department of Physics, Ohio State University, Columbus, OH 43210}


\date{\today}

\begin{abstract}

We investigate the ground state of interacting spin-$\half$ fermions (3D)
at a finite density ($\rho \sim \kf^3$) in the presence of a uniform
non-Abelian gauge field. The gauge field configuration (GFC) described by a
vector $\blam \equiv (\lambda_x, \lambda_y, \lambda_z)$, whose
magnitude $\lambda$ determines the gauge coupling strength, generates
a generalized Rashba spin-orbit interaction. For a weak 
attractive interaction in the singlet channel described by a small negative scattering length
$(\kf |\as| \lesssim 1)$, the ground state in the absence of the gauge
field ($\lambda=0$) is a BCS (Bardeen-Cooper-Schrieffer) superfluid
with large overlapping pairs. With increasing gauge coupling strength,
a non-Abelian gauge field engenders a crossover of this BCS ground
state to a BEC (Bose-Einstein condensate) ground state of bosons even
with a weak attractive interaction that fails to produce a two-body bound
state in free vacuum. For large gauge couplings $(\lambda/\kf \gg 1)$,
the BEC attained is a condensate of bosons whose properties
are solely determined by the gauge field (and not by the scattering
length so long as it is non-zero) -- we call these bosons
``rashbons''. In the absence of interactions ($\as = 0^-$), the shape of the Fermi
surface of the system undergoes a topological transition at a
critical gauge coupling $\lambda_T$. For high symmetry gauge field
configurations we show that the crossover from the BCS superfluid to
the rashbon BEC occurs in the regime of $\lambda$ near
$\lambda_T$. In the context of cold atomic systems, this work makes an interesting suggestion of obtaining BCS-BEC crossover through a route other than tuning the interaction
between the fermions.

\end{abstract}


\maketitle

\section{Introduction}
\mylabel{sec:Intro}

Recent experimental progress in the generation of synthetic gauge
fields\cite{Lin2009A, Lin2009B, Lin2011} has enhanced the
possibilities of controlled experimental studies of outstanding problems
of quantum condensed matter and even of high energy
physics\cite{Maeda2009} using cold atomic systems. Many theoretical
works\cite{Jaksch2003,Osterloh2005,Ruseckas2005,Gerbier2010} have
explored the possibilities of generating both Abelian and non-Abeilan
gauge fields. The experimental work with synthetic gauge fields have
been with bosonic $^{87}$Rb atoms. In a recent commentary\footnote{See
  commentary on reference \myonlinecite{Lin2011} by T.-L. Ho at
  http://www.condmatjournalclub.org} , the investigation of fermions
in synthetic non-Abelian gauge fields has been identified as a key
research direction.

The study of interacting fermions in 3D space with controlled
interactions has been one of the key successes of cold atoms
research.\cite{Ketterle2008,Bloch2008,Giorgini2008} A particular
example is the problem of the crossover from a BCS ground state with
large overlapping pairs to a BEC of tightly bound bosonic pairs with
increasing strength of attractive interactions -- a phenomenon that
was suggested many years
earlier.\cite{Eagles1969,Leggett1980,Leggett2006} The superfluid
transition temperature on the BCS side is determined by the superfluid
energy gap, while on the BEC side the transition temperature is
determined by the condensation temperature of the tightly bound
bosonic pairs of fermions.\cite{Nozieres1985} A review of BCS-BEC
crossover that is particularly useful in the context of this paper may
be found in reference~\myonlinecite{Randeria1995}.

An interesting question is regarding the fate of interacting fermions
in the presence of a non-Abelian gauge field. Motivated by the fact
that even a spatially uniform non-Abelian gauge field produces
interesting physical effects for bosons,\cite{Ho2010,Wang2010,Lin2011}
we focus on interacting fermions in uniform non-Abelian gauge fields.

In a recent paper,\cite{Vyasanakere2011} two of us investigated how a
uniform non-Abelian gauge field influences the bound state of two
spin-$\half$ fermions interacting via a contact attraction in the
singlet channel characterized by a s-wave scattering length $\as$.  The type
of uniform non-Abelian gauge field considered in that work leads to a
generalized Rashba spin orbit interaction. A key finding of that work
is that for high symmetry gauge field configurations (more precisely
defined in the next section), a two-body bound state exists for
{\em any} scattering length however small and negative. The study
suggested that the BCS-BEC crossover is drastically affected by the
presence of a non-Abelian gauge field.

Here we pursue the ideas of reference \myonlinecite{Vyasanakere2011}
through a study of the ground state of a finite density of interacting
fermions in a non-Abelian gauge field by means of mean field theory.
We show that increasing the strength of a non-Abelian gauge field
produces a crossover from a BCS superfluid (which is the ground state
in the absence of the gauge field) to a BEC of bosons at a fixed
interaction (fixed scattering length $\as$) however small and
negative. Further, {\em the bosons that condense to form the BEC at
  large gauge couplings are tightly bound pairs of fermions whose
  properties are determined solely by the non-Abelian gauge field} --
we have ventured to call these bosons ``rashbons''. For a given
attractive interaction (fixed $\as$), therefore, the crossover takes
place the standard BCS superfluid state to a rashbon BEC. There is an
additional feature of the crossover that is particularly
noteworthy. The Fermi surface of the non-interacting system ($\as =
0^-$) undergoes a transition in its topology with increasing
gauge coupling strength. We show that for high symmetry gauge field
configurations, the crossover regime of gauge couplings in the
presence of interactions overlaps with the regime of topological
transition of the non-interacting Fermi surface. In a sense this
provides a ``geometrical'' view of the crossover.

The statement of the problem we address
and a detailed summary of our results are given in
Section~\ref{sec:ProbStatementAndSummary}. The mean field formulation is
detailed in Section~\ref{sec:MFT}. Section~\ref{sec:Results} describes
the results. The paper is concluded with a discussion in
Section~\ref{sec:Discussion}. We recommend the reading of Section~\ref{sec:ProbStatementAndSummary} and Section~\ref{sec:Discussion} to obtain a physical picture of our results.

\section{Question Addressed and Summary of Results}
\mylabel{sec:ProbStatementAndSummary}

In units where the mass of the fermions and Planck's constant are set
to unity, the hamiltonian of the fermions moving in a uniform
non-Abelian gauge field is
\begin{equation}
{\cal H}_{GF} = \int \D{^3 \br} \, \Psi^\dagger(\br) \left[ \half (\bp \bOne - \bA^\mu \tau^\mu) \cdot (\bp \bOne - \bA^\mu \tau^\mu)  \right] \Psi(\br), \mylabel{eqn:HGF}
\end{equation}
where $\Psi(\br) = \{\psi_{\sigma}(\br)\}, \sigma= \uparrow,
\downarrow$ are fermion operators, $\bp$ is the momentum, $\bA^\mu
\equiv A_i^\mu \be_i$, are uniform gauge fields, $\tau^\mu$ ($\mu
=x,y,z$) are Pauli matrices and $\be_i$'s are the unit vectors in the $i$-th direction, $i=x,y,z$. As in \myonlinecite{Vyasanakere2011}, we
specialize to $A_i^\mu= \lambda_i \delta_i^\mu$ leading to a hamiltonian
with a generalized Rashba spin-orbit interaction
\beq
{\cal H}_R = \int \D{^3 \br} \, \Psi^\dagger(\br) \left( \frac{\bp^2}{2} \bOne - \bp_\lambda \cdot \btau \right) \Psi(\br), \mylabel{eqn:HRashba}
\eeq
where $\bp_\lambda = \sum_i p_i \lambda_i \be_i$. The vector $\blam =
\lambda \hat{\blam} = \sum_i \lambda_i \be_i$ describes a gauge field
configuration (GFC) space as depicted in \fig{fig:GFCSpace}; we call
$\lambda = |\blam|$ as the gauge coupling strength. High symmetry GFCs
are important and have been classified in
\myonlinecite{Vyasanakere2011} as prolate, spherical and oblate. Of
particular interest are the configurations shown in \fig{fig:GFCSpace}
called extreme prolate (EP), spherical(S) and extreme oblate(EO).

\begin{figure}
\centerline{\includegraphics[width=\myfigwidth]{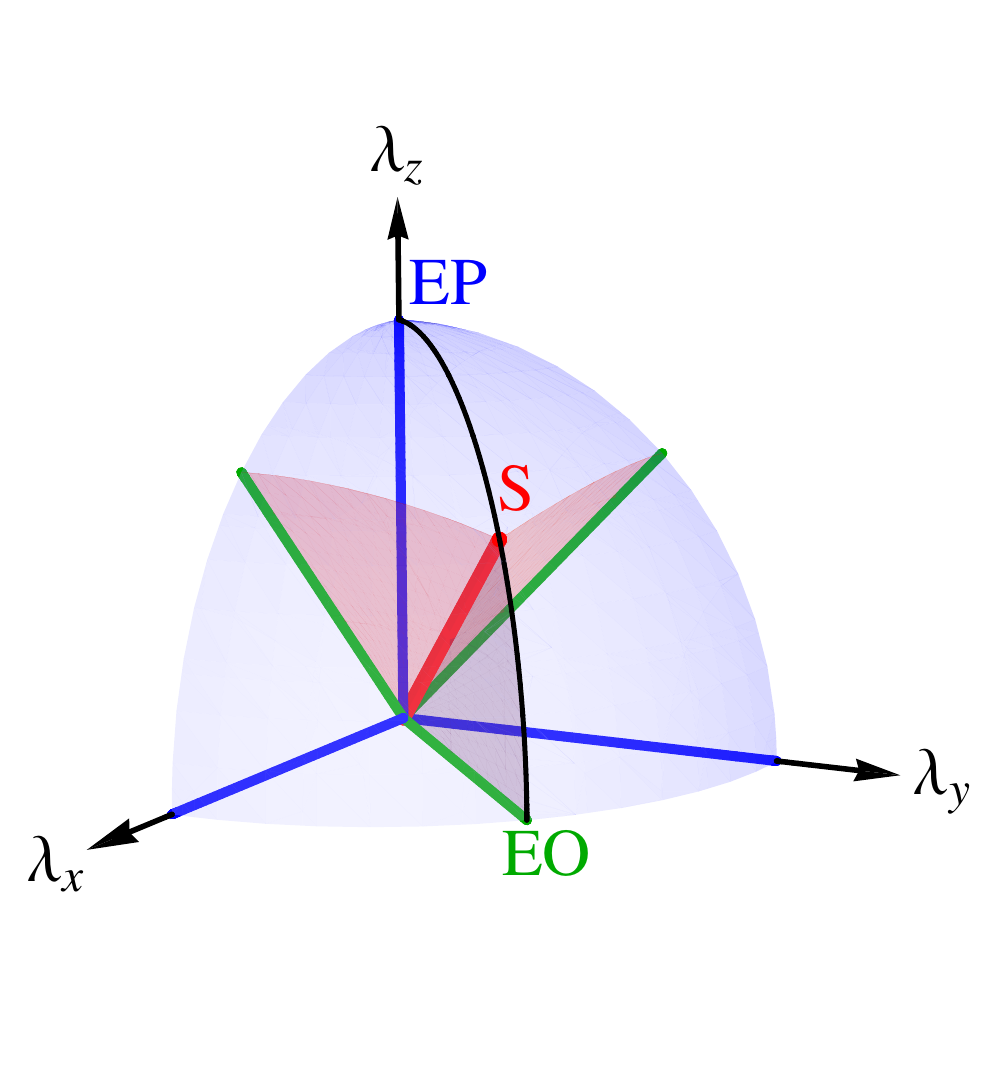}}
\caption{(Color online) Gauge field configuration (GFC) space. The non-Abelian gauge
  field of \protect{\eqn{eqn:HRashba}} is described by a vector $\blam
  = (\lambda_x, \lambda_y, \lambda_z) = \lambda \hat{\blam}$, where
  $\lambda = |\blam|$ is the gauge coupling strength and $\hat{\blam}$
  is a unit vector. High symmetry GFCs such as extreme prolate (EP,
  $\hat{\blam} = (0,0,1)$), spherical (S,
  $\hat{\blam}=\frac{1}{\sqrt{3}}(1,1,1)$) and extreme oblate (EO,
  $\hat{\blam}=\frac{1}{\sqrt{2}}(1,1,0)$) are as shown. 
}
\mylabel{fig:GFCSpace}
\end{figure}

The one-particle states of ${\cal H}_R$ are 
\beq
\ket{\bk \alpha} = \ket{\bk} \otimes \ket{\alpha \hat{\bk}_\lambda} \mylabel{eqn:HelicityEigenStates}
\eeq
that disperse as
\beq
\varepsilon_{\bk \alpha} = \frac{k^2}{2} - \alpha |\bk_\lambda| \mylabel{eqn:RashbaDispersion}
\eeq
where $\bk$-the momentum, and $\alpha = \pm 1$-the eigenvalues of the
helicity operator $\hat{\bp}_\lambda \cdot \btau$, are the good
quantum numbers. The quantity $\bk_\lambda$ is defined analogously with
$\bp_\lambda$ in \eqn{eqn:HRashba}. For any $\blam$, the two
helicity states for a given $\bk$ are degenerate only at $\bk ={\bf
  0}$.

The interaction between the fermions is described by contact attraction in the singlet channel
\beq
{\cal H}_{\upsilon} =      \upsilon     \int \D{^3 \br}\, \psi^\dagger_\uparrow(\br) \psi^\dagger_{\downarrow}(\br) \psi_{\downarrow}(\br) \psi_{\uparrow}(\br) \mylabel{eqn:ContactAttraction}.
\eeq
The endemic ultraviolet divergence of the theory described by the hamiltonian 
\beq
{\cal H} = {\cal H}_R + {\cal H}_\upsilon \mylabel{eqn:FullHamiltonian}
\eeq
is handled \cite{Braaten2008} by introducing an ultraviolet momentum cutoff
$\Lambda$. This entails characterization of the attraction by a
physical parameter that describes the low energy scattering properties
while making the parameter $\upsilon$ depend on $\Lambda$. More
precisely,
\beq
\frac{1}{\upsilon} + \Lambda = \frac{1}{4 \pi \as}
\eeq
where $\as$ is the $s$-wave scattering length in free vacuum, i.~e.,
when the gauge field is absent ($\lambda=0$). In free vacuum (3D) only an attraction
larger than a critical strength can produce a two-particle bound
state. This is embodied in the fact that for $\as < 0$ (BCS side)
there is no two-body bound state; a bound state develops only as $\as
\rightarrow -\infty$, or $\frac{1}{\as} \rightarrow 0^-$
(resonance). For $\as >0$ (BEC side) a bound state is obtained with a
binding energy $\Eb = \frac{1}{\as^2}$.

\begin{table*}
\centerline{
\renewcommand{\arraystretch}{2.5}
\begin{tabular}{||c||c||c|c|p{3.55truecm}||c|c|p{3.55truecm}||c|c|c||} \hline
\hline 
\multirow{3}*{GFC} & \multirow{3}{*}{$\asc$} & \multicolumn{3}{|c||}{$\as < 0$} & \multicolumn{3}{|c||}{Resonance} & \multicolumn{3}{|c||}{$\as > 0$} \\ 
                 &   & \multicolumn{3}{|c||}{$\lambda |\as| \ll 1$} &  \multicolumn{3}{|c||}{$1/(\lambda \as) = 0$} &  \multicolumn{3}{|c||}{$\lambda \as \ll 1$} \\ \cline{3-11}
&  & $\Eb$ & $\etaT$ & \centerline{Spin Structure} & $\Eb$ & $\etaT$ &\centerline{Spin Structure}& $\Eb$ & $\etaT$ & Spin Structure \\ \hline 
\hline
EP & $-\infty$ & \multicolumn{3}{|c||}{No bound state} & 0 & $\;\;\displaystyle{\frac{1}{2}}\;\;$  & \centerline{Bi-axial nematic (BW)} & $\displaystyle{\frac{1}{\as^2}}$ & 0 & singlet \\
\hline \hline
S & $0^-$ & $\displaystyle{\frac{\lambda^4 \as^2}{3}}$ & $\;\;\displaystyle{\frac{1}{2}}\;\;$ & \centerline{Spherical} & $\displaystyle{\frac{\lambda^2}{3}}$ & $\displaystyle{\frac{1}{4}}$ &\centerline{Spherical} & $ \displaystyle{\frac{1}{\as^2} + \frac{2\lambda^2}{3}}$ & 0 & singlet \\ \hline \hline
EO & $0^-$ & $\displaystyle{\frac{2}{e^2}e^{-\frac{2\sqrt{2}}{\lambda |\as|}} }$ & $\;\;\displaystyle{\frac{1}{2}}\;\;$ & \centerline{Uni-axial nematic (ABM)} & $\displaystyle{0.22 \lambda^2}$ & 0.28 & \centerline{Uni-axial nematic (ABM)} &  $ \displaystyle{\frac{1}{\as^2} + \frac{\lambda^2}{2}}$ & 0 & singlet \\ \hline \hline
\end{tabular}
}
\caption{Summary of the two body problem\cite{Vyasanakere2011} in high
  symmetry GFCs $(\lambda > 0)$. $E_b$ is the binding energy and $\etaT$ is the
  triplet fraction (not reported in
  reference~\myonlinecite{Vyasanakere2011}). The bi-axial spin nematic
  structure is similar to the BW (Balian-Werthamer) or B-phase of
  $^3$He, and the uni-axial nematic structure to that of the ABM
  (Anderson-Brinkman-Morel) or A-phase of $^3$He. The values of $\Eb$
  and $\etaT$ at resonance, which correspond to the properties of the
  rashbon (for S and EO cases), are exact results, while others are
  asymptotic values.
}
\mylabel{tab:TwoBodyInfo}
\end{table*}

A uniform non-Abelian gauge field brings about remarkable changes in
the two body problem as shown in reference
\myonlinecite{Vyasanakere2011}. Most vividly, for high symmetry GFCs
such as S and EO, the critical scattering length $\asc$ required for
the formation of a bound state vanishes, i.~e., there is a two body
bound state for {\em any} scattering length however small and negative
(deep BCS side). The size of the binding energy of the bound state
depends on the GFC. On the BCS side, the binding energy has an
exponential dependence on the scattering length and $\lambda$ for the EO GFC while
for the S GFC this dependence is algebraic. Another
interesting aspect that emerges is the symmetry of the bound-state
wave function. In a non-Abelian gauge field the normalized bound-state wave
function is made up of spatially symmetric singlet and spatially
antisymmetric triplet pieces
\beq
\ket{\psi_b} = \ket{\psi_s} + \ket{\psi_t}. \mylabel{eqn:TripletSplit}
\eeq
Time reversal symmetry of the hamiltonian is preserved and this
two-body-bound-state wave function picks up a nematic spin structure
consistent with the symmetry of the GFC.
 \footnote{\label{NematicFN} \protect{ The quadrupole operator is defined as $Q^{\alpha \beta} = \half \left( {\cal S}^\alpha {\cal S}^\beta + {\cal S}^{\beta} {\cal S}^{\alpha}\right) -\frac{\mean{{\cal S}^2}}{3} \delta^{\alpha \beta}$ where ${\cal S}^\alpha$ are
    spin operators.  A nematic state has $\mean{{\cal S}^{\alpha}} =0$, while $\mean{Q^{\alpha \beta}} \ne 0$.  In the present context, the
  singlet piece of the two body wave function does not contribute to the quadrupole
  moment, while the triplet wave function has $\mean{{\cal S}^\alpha} = 0$, but $\mean{Q^{\alpha \beta}}
  \ne 0$
}
} The triplet content of the
wave function which also measures the ``amount of nematicity'' is
characterized by a parameter which we call the triplet fraction
\beq
\etaT = \braket{\psi_t}{\psi_t}. \mylabel{eqn:etaT}
\eeq
The binding energy $\Eb$, the triplet fraction $\etaT$ and
the spin symmetry of the two-body wave function for different GFCs are
summarized \footnote{It is to be noted that the result for the EO case
  presented in \myonlinecite{Vyasanakere2011} has a minor error of a
  factor of 2. The correct results are quoted here; see also Section \Rmnum{3} B of
  e-print arXiv:1101.0411v2} in Table~\ref{tab:TwoBodyInfo}. The gist
of reference \myonlinecite{Vyasanakere2011} is that high symmetry GFCs
induce high degeneracy in the low energy (infrared) one particle
density of states and this promotes bound state
formation. Colloquially, high symmetry GFCs are ``attractive
interaction amplifiers''. An aspect of the two body problem that
  is important in the discussion below is that the physics of the two
  body problem in the presence of the gauge field ($\lambda > 0$) is
  completely determined by the dimensionless parameter $\lambda
  \as$. All aspects of the solution depends only on $\lambda \as$ when
  length and energy are respectively measured in units of
  $\lambda^{-1}$ and $\lambda^2$. This is true for {\em any GFC except EP GFC} as is evident from Table.~\ref{tab:TwoBodyInfo}

\begin{figure*}
\centerline{
\begin{minipage}{5.0truecm}
\centerline{\includegraphics[width=4.25truecm]{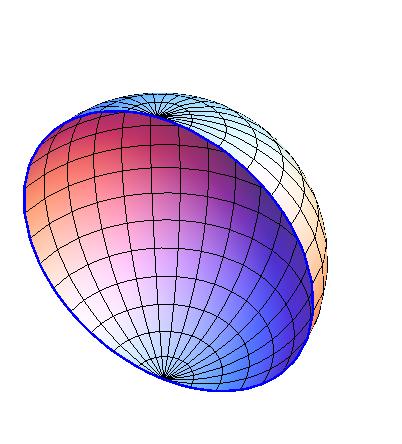}}
\centerline{(a) $\lambda = 0$} 
\end{minipage}
\begin{minipage}{5.0truecm}
\centerline{\includegraphics[width=5.0truecm]{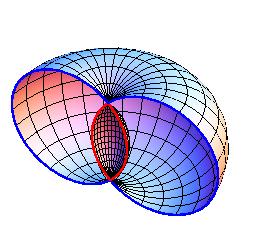}}
\centerline{(b) $0 < \lambda < \lambda_T$} 
\end{minipage}
\begin{minipage}{5.0truecm}
\centerline{\includegraphics[width=5.0truecm]{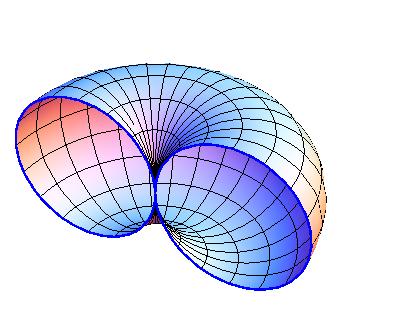}}
\centerline{(c) $\lambda = \lambda_T$} 
\end{minipage}
\begin{minipage}{5.0truecm}
\centerline{\includegraphics[width=5.0truecm]{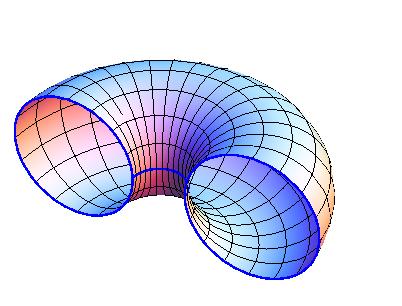}}
\centerline{(d) $\lambda > \lambda_T$} 
\end{minipage}
}
\caption{(Color online) Fermi surface topology transition (FSTT) with increasing
  gauge coupling strength for EO GFC. (a) Two overlapping spherical Fermi
  surfaces in the absence of a gauge field (b) A gauge coupling
  smaller than $\lambda_T$: The union of the $+$ and $-$ Fermi
  surfaces forms a spindle torus. The apple of the spindle torus is the $+$
  helicity Fermi surface which is shown with a blue border, and the lemon of
  the spindle torus is the $-$ helicity Fermi surface which is shown with a red border. (c) The gauge
  coupling that obtains the FSTT. The $+$ helicity Fermi surface is a
  horn torus, while the $-$ helicity Fermi surface vanishes. (d) For a
  gauge coupling larger than $\lambda_T$, there is only the $+$
  helicity Fermi surface which is a ring torus. Note that all the figures show only a sectioned half of the Fermi surfaces.}
\mylabel{fig:EOTopologicalTransition}
\end{figure*}

In this paper we investigate the system described by the hamiltonian
of \eqn{eqn:FullHamiltonian} at {\em finite} density $\rho$ of the
fermions. The finite density of particles introduces an additional
energy scale which can be conveniently taken to be the Fermi energy
$\Ef$ (and an associated Fermi wave vector $\kf$) in the {\em absence}
of the gauge field ($\lambda =0$)
\beq
\Ef = \frac{\kf^2}{2} = \half {(3 \pi^2 \rho)}^{2/3}. \mylabel{eqn:Efdef}
\eeq
At finite densities of fermions, therefore, the ground state of the
system (zero temperature) of \eqn{eqn:FullHamiltonian} is determined
by the dimensionless parameters $\kf \as$, $\lambda/\kf$, and the
direction $\hat{\blam}$ in GFC space. In this paper we work at fixed
density and study the evolution of the ground state of the system with
$\lambda/\kf$ and $\kf \as$ for various $\hat{\blam}$s corresponding
to high symmetry GFCs.

The possibility of interesting physics in this system is suggested by
the following observation which provides the motivation for this
work. Consider a system of {\em non-interacting} (NI) fermions (${\cal
  H}_\upsilon = 0 $ in \eqn{eqn:FullHamiltonian}). In the absence of a
gauge field ($\lambda =0$), the ground state has a chemical potential
$\Ef$ and is described by two identical filled Fermi seas bounded by
spherical Fermi surfaces of radius $k_F$ -- one each for $\uparrow$
and $\downarrow$ spins. In the presence of the gauge field $(\lambda
\ne 0)$, the helicity $\alpha$ is the good quantum number along with
momentum (see \eqn{eqn:HelicityEigenStates}), and hence the ground
state will be two Fermi seas, one for each helicity. The chemical
potential\footnote{Throughout the paper, the chemical potential is
  referred to the bottom of the $+$ helicity band.} now depends on
$\lambda$ through a function $\muNI(\lambda)$ that is determined by
$\hat{\blam}$. Both of the Fermi seas are generically non-spherical
with a shape determined by $\hat{\blam}$.  Since the one particle
states with opposite helicities but with same momentum are
non-degenerate (see \eqn{eqn:RashbaDispersion}), the Fermi surfaces of
different helicities are not identical and evolve differently with
increasing $\lambda$ (at a fixed density $\rho$). The most interesting
aspect is that since the $+$ helicity state is lower in energy than the $-$
helicity one for all momenta, upon increasing $\lambda$, the volume
enclosed by the $+$ helicity Fermi surface increases at the expense of
that of the $-$ helicity Fermi surface. Matters come to a head at a
critical gauge coupling $\lambda_T$ (which depends on $\hat{\blam}$)
where the $-$ helicity Fermi sea ceases to exist since the chemical
potential $\muNI(\lambda)$ falls below the bottom of the $-$ helicity
band. Thus, for $\lambda \geq \lambda_T$ the ground state is a Fermi
sea of only $+$ helicity. This is illustrated for the EO GFC in
\fig{fig:EOTopologicalTransition}. The values of $\lambda_T$
determined by the density of particles for various high symmetry GFCs
are given in Table~\ref{tab:SummaryOfResults}; it is to be noted that
in all cases $\lambda_T$ is of order $\kf$. Another aspect to be noted is
that there is a {\em change in the topology} of the $+$ helicity Fermi
surface at $\lambda_T$. We call this the Fermi surface topology
transition (FSTT) -- hence the subscript $T$ in $\lambda_T$. For
example, in the EO case, the genus of the $+$ helicity Fermi surface
changes from zero (homeomorphic to a sphere) to unity (homeomorphic to
a torus) at $\lambda_T$ as illustrated in
\fig{fig:EOTopologicalTransition}.

\begin{table*}
\centerline{
\renewcommand{\arraystretch}{2.5}
\begin{tabular}{||c||c||c|c|c||c|c|c||c||} \hline
\hline 
\multirow{3}*{GFC} & \multirow{3}{*}{$\lambda_T$} & \multicolumn{3}{|c||}{$\lambda \ll \lambda_T$} & \multicolumn{3}{|c||}{$\lambda \gg \lambda_T$} & Crossover to   \\ 
                 &   & \multicolumn{3}{|c||}{Before FSTT} &  \multicolumn{3}{|c||}{After FSTT}  &  rashbon-BEC    \\ \cline{3-8}
&  & $\mu$ & $\etaT$ & Spin Structure & $\mu$ & $\etaT$ & Spin Structure &   as $\lambda \rightarrow \infty$? \\
\hline \hline
EP & $\kf$ & $\approx \Ef $ & $\propto \lambda^2$ & Bi-axial nematic &$\approx \Ef$  &  $\displaystyle{\frac{1}{2}}$  & Bi-axial nematic & No \\
\hline \hline
S & $\frac{\sqrt{3}}{2^{(2/3)}} \kf$ & $\approx \muNI(\lambda)$ (\eqn{eqn:SmuNI}) & $\propto \lambda^2$ & Spherical & $\approx -\displaystyle{\frac{\lambda^2}{6}}$  &  $\approx \displaystyle{\frac{1}{4}}$  & Spherical & Yes \\
\hline \hline
EO & $ \left(\frac{8\sqrt{2}}{3\pi}\right)^{\frac{1}{3}} \kf$ & $\approx\muNI(\lambda)$ (\eqn{eqn:EOmuNI})  & $\propto \lambda^2$ & Uni-axial nematic &$\approx -0.11 \lambda^2$  &  $\approx 0.28$   & Uni-axial nematic & Yes \\ \hline \hline
\end{tabular}
}
\caption{Summary of the properties of the superfluid ground state for
  two regimes of the gauge coupling strength $\lambda$. The results
  shown here are for a weak attractive interaction ($\as < 0, \kf|\as|
  \ll 1$). $\mu$ is the ground state chemical potential and $\etaT$ is
  the triplet content of the pair wave function.  Well before the FSTT
  ($\lambda \ll \lambda_T$), the ground state is a BCS superfluid.  In
  the regime ($\lambda \gg \lambda_T$ ), for the S and EO GFCs, the
chemical potential is close to half of the rashbon energy (see
Table.~\protect{\ref{tab:TwoBodyInfo}}) and the pair wave function attains the
same triplet content as that of the rashbon, clearly indicating a
crossover from a BCS superfluid to a rashbon condensate. Any GFC {\em except} EP produces a BCS - rashbon BEC crossover.}
\mylabel{tab:SummaryOfResults}
\end{table*}

What happens when the interaction ${\cal H}_\upsilon$
(\eqn{eqn:ContactAttraction}) is turned on? Consider an interaction
with a small negative scattering length ($k_F |\as| \ll 0$, deep BCS
side). For $\lambda \ll \lambda_T$, the ground state is a superfluid
with overlapping pairs with an exponentially small excitation gap. The
chemical potential of this state is nearly unaffected and is
$\muNI(\lambda)$, that of the non-interacting system in a gauge
field. The only qualitative difference from the usual $s$-wave BCS
state is that the pair wave function now has a small triplet content
and associated spin nematicity induced by the gauge field. This
picture changes drastically in the case of high symmetry GFCs (such as
S and EO) when the gauge coupling strength $\lambda$ is tuned past
$\lambda_T$. The key finding of this paper is that {\em for high-symmetry GFCs, a BEC of tightly bound pairs is obtained for $\lambda
  \gg \lambda_T$ even with a small negative scattering length.} In
other words, one can engineer a BCS-BEC crossover with a high-symmetry
GFC by increasing the gauge coupling strength even with a very weak
attractive interaction that is unable to produce a two-body bound-state in free vacuum. This result arises from the fact that for
$\lambda \gg \lambda_T$, the size of the {\em two-body} bound-state
wave function (see Table~\ref{tab:TwoBodyInfo}) becomes smaller than
the inter-particle spacing. The fermions therefore form tightly bound pairs which then Bose condense in the
zero center of mass momentum state. As is evident from the discussion,
the physics of these results owes to the character of high-symmetry
GFCs to act as attractive-interaction-amplifiers. Indeed, as
$\lambda/\lambda_T \rightarrow \infty$ the chemical potential tends to
that determined by the two-particle bound-state energy. Since $\lambda |\as| \rightarrow \infty$  (fixed $\as$),
the nature of the {\em two body bound state obtained is identical to that obtained with a resonant scattering length  in the presence of the gauge field} as tabulated in
Table~\ref{tab:TwoBodyInfo}. The properties of this bosonic bound
state of two fermions is {\em determined solely by the Rashba gauge
  field}; we call these emergent bosons as ``rashbons'' (see
Sec.~\ref{sec:SResults} and second para of Sec.~\ref{sec:Discussion} for details). The BEC that is obtained for
$\lambda \gg \lambda_T$ is a rashbon condensate.  The results for
various GFCs are tabulated in Table~\ref{tab:SummaryOfResults} which
is a summary of this paper.

In the remaining sections we illustrate these conclusions by
a mean field theory of the superfluid ground state of this
interacting fermion system. Mean field theory is known to give a
qualitatively correct description for the superfluid ground
state.\cite{Diener2008}

\section{Mean Field Theory}
\mylabel{sec:MFT}

We now describe the details of the mean field analysis of the
superfluid ground state of fermions in a non-Abelian gauge field. This
analysis involves certain straightforward manipulations beyond the standard
formulation,\cite{Randeria1995} and hence presented in detail. We note that the present analysis can
treat {\em any} GFC. Results for specific GFCs of interest will be
presented in the next section.

To perform a mean-field analysis of the superfluid ground state we
recast the interaction term ${\cal H}_\upsilon$ in a convenient form
\beq
{\cal H}_{\upsilon} = \frac{\upsilon}{2} \int \D{^3\br} \, S^\dagger(\br) S(\br), \mylabel{eqn:HvSinglet}
\eeq
where $S^\dagger(\br)$ is the singlet creation operator
\beq
S^{\dagger}(\br) = \frac{1}{\sqrt{2}} \left(\psi^\dagger_\uparrow(\br) \psi^\dagger_{\downarrow}(\br) - \psi^\dagger_{\downarrow}(\br) \psi^\dagger_{\uparrow}(\br) \right) \mylabel{eqn:SingletDef}.
\eeq
The interaction  in terms of the singlet operators can be cast in momentum space 
\beq
{\cal H}_\upsilon = \frac{\upsilon}{2 V} \sum_{\bq} S^\dagger(\bq) S(\bq),
\eeq
where $V$ is the volume of the system and the mean-field ansatz corresponds to taking
\beq
\mean{S(\bq)} = \mean{S} \delta_{\bq,\bzero},
\eeq
where $\mean{S} = \mean{S(\bzero)}$ with
\bea
S^\dagger(\bzero) = \frac{1}{\sqrt{2}} \sum_{\bk} \left(c^\dagger_{\bk \uparrow} c^\dagger_{-\bk \downarrow} - c^\dagger_{\bk \downarrow} c^\dagger_{-\bk \uparrow} \right). \mylabel{eqn:Stmp}
\eea
The fermion operators $c^\dagger_{\bk \sigma}$ are defined by
\beq
c^\dagger_{\bk \sigma} = \frac{1}{\sqrt{V}} \int \D{^3{\br}} \, e^{-i \bk \cdot \br} \psi^\dagger_{\sigma}(\br).
\eeq
It is now convenient to sum {\em only over half} of the allowed $\bk$ values in \eqn{eqn:Stmp}
\beq\mylabel{eqn:HalfSum}
S(\bzero) = \sqrt{2} \sum_{\bk}'  \left(c^\dagger_{\bk \uparrow} c^\dagger_{-\bk \downarrow} - c^\dagger_{\bk \downarrow} c^\dagger_{-\bk \uparrow} \right)
\eeq
as is here and henceforth indicated by the prime over the summation
symbol. The advantage of this exercise is that the operator
$S(\bzero)$ can be written in the helicity basis as
\beq
S(\bzero) = \sqrt{2} \sum_{\bk \alpha}' \alpha c^\dagger_{\bk \alpha} c^\dagger_{-\bk \alpha}
\eeq
Introducing a chemical potential $\mu$, we obtain the mean-field Hamiltonian as
\beq\mylabel{eqn:MeanFieldHamTmp}
\begin{split}
{\cal H}_{MF} & = \sum_{\bk \alpha} \xi_{\bk \alpha} c^\dagger_{\bk \alpha} c_{\bk \alpha} + \Delta \sum_{\bk \alpha}' \alpha c^{\dagger}_{\bk \alpha} c^\dagger_{-\bk \alpha} \\
&  +  \Delta \sum_{\bk \alpha}' \alpha c_{-\bk \alpha} c_{\bk \alpha}  - \frac{V \Delta^2}{\upsilon} 
\end{split}
\eeq
where $c^\dagger_{\bk \alpha}$ are electron operators associated with
the one-particle helicity eigenstate (\eqn{eqn:HelicityEigenStates}),
$\Delta = \frac{\upsilon \mean{S}}{\sqrt{2} V}$ is the order parameter
(taken to be real), and $\xi_{\bk \alpha} = \tilde{\varepsilon}_{\bk \alpha} -
\mu$. Here $\tilde{\varepsilon}_{\bk \alpha}$ is $\varepsilon_{\bk \alpha}$ referred to the bottom of the $+$ helicity band. Noting inversion symmetry, $\xi_{-\bk \alpha}=\xi_{\bk
  \alpha}$, eqn.~\prn{eqn:MeanFieldHamTmp} can now be recast as
\beq\mylabel{eqn:MeanFieldHam}
\begin{split}
{\cal H}_{MF} & = \sum_{\bk \alpha}' \left(
\begin{array}{cc}
c^\dagger_{\bk \alpha} & c_{-\bk \alpha} 
\end{array}
\right)
\left[
\begin{array}{cc}
\xi_{\bk \alpha} & \alpha \Delta \\
\alpha \Delta & -\xi_{\bk \alpha}
\end{array} 
\right] 
\left(
\begin{array}{c}
c_{\bk \alpha} \\
c^\dagger_{-\bk \alpha}
\end{array}
\right)
\\
& + \sum_{\bk \alpha}' \xi_{\bk \alpha}    - \frac{V \Delta^2}{\upsilon} 
\end{split}
\eeq
which now has the standard form except for the fact that the summation
over $\bk$ is carried out only over half of the momentum space and a
sum over the two helicities is taken.

The hamiltonian in \eqn{eqn:MeanFieldHam} can now be diagonalized in
terms of the Bogoliubov quasiparticle operators as
\beq \mylabel{eqn:MFdiagonalized}
\begin{split}
{\cal H}_{MF} & = \sum_{\bk \alpha}' E_{\bk \alpha} \left(\gamma^\dagger_{\bk \alpha 1} \gamma_{\bk \alpha 1} + \gamma^\dagger_{\bk \alpha 2} \gamma_{\bk \alpha 2} \right) \\
& + \sum_{\bk \alpha}' \left( \xi_{\bk \alpha} - E_{\bk \alpha} \right)  - \frac{V \Delta^2}{\upsilon} 
\end{split}
\eeq
where $E_{\bk \alpha} = \sqrt{\xi_{\bk \alpha}^2 + \Delta^2} $
($\Delta$ is also the excitation gap), and
\beq\mylabel{eqn:Bogoliubov}
\begin{split}
\gamma_{\bk \alpha 1} & = u_{\bk \alpha} c_{\bk \alpha} - \alpha v_{\bk} c^\dagger_{-\bk \alpha} \\
\gamma^\dagger_{\bk \alpha 2} & = \alpha v_{\bk \alpha} c_{\bk \alpha} + u_{\bk} c^\dagger_{-\bk \alpha}
\end{split}
\eeq
with
\beq
u^2_{\bk \alpha} = \half \left(1 + \frac{\xi_{\bk \alpha}}{E_{\bk \alpha}} \right),  \;\;\; v^2_{\bk \alpha} = \half \left(1 - \frac{\xi_{\bk \alpha}}{E_{\bk \alpha}} \right) .
\eeq
A standard analysis now leads to the gap equation
\beq
-\frac{1}{\upsilon} = \sum_{\bk \alpha}' \frac{1}{2 E_{\bk \alpha}}.
\eeq
Noting the inversion symmetry of the problem and using the
 renormalization of the interaction, the gap equation becomes
\beq
-\frac{1}{4 \pi \as} = \frac{1}{2V} \sum_{\bk \alpha} \left( \frac{1}{2E_{\bk \alpha}} - \frac{1}{k^2} \right) \mylabel{eqn:Gap}.
\eeq
The number equation is
\beq
\rho = \frac{1}{V} \sum_{\bk \alpha} \half{\left(1-\frac{\xi_{\bk \alpha}}{E_{\bk \alpha}}\right)}. \mylabel{eqn:Number}
\eeq
The solution of \eqn{eqn:Gap} along with the number equation
\eqn{eqn:Number}, determines the chemical potential $\mu$ and the gap parameter $\Delta$ in the ground state.

The ground state $\ket{\Psi_{G}}$ of the system is given by 
\beq\mylabel{eqn:GroundState}
\ket{\Psi_{G}} = \prod_{\bk\alpha}' (u_{\bk\alpha} + \alpha v_{\bk \alpha} c^\dagger_{\bk \alpha}c^\dagger_{-\bk \alpha}) \ket{0}
\eeq
where $\ket{0}$ is the fermion vacuum. This can be (up to a
normalization) be re-written as
\beq
\ket{\Psi_{G}} = e^{P^\dagger}  \ket{0}
\eeq
where $P^\dagger$ is the pair creation operator given by
\beq\mylabel{eqn:Pair}
P^\dagger = \sum_{\bk \alpha}' \alpha \phi_{\bk \alpha} c^\dagger_{\bk \alpha} c^\dagger_{-\bk \alpha} 
\eeq
where $\displaystyle{\phi_{\bk \alpha} = \frac{v_{\bk \alpha}}{u_{\bk \alpha}}}$. 
The singlet and triplet parts of the pair can be extracted by noting that 
\beq\mylabel{eqn:PairST}
\begin{split}
P^\dagger &= \underbrace{ \sum_{\bk \alpha}' \phi_s(\bk) \left(c^\dagger_{\bk+} c^\dagger_{-\bk+} - c^\dagger_{\bk-} c^\dagger_{-\bk-} \right)}_{\mbox{singlet}} \\
& + \underbrace{ \sum_{\bk \alpha}' \phi_t(\bk) \left(c^\dagger_{\bk+} c^\dagger_{-\bk+} + c^\dagger_{\bk-} c^\dagger_{-\bk-} \right)}_{\mbox{triplet}} 
\end{split}
\eeq
with
\beq\mylabel{eqn:PhisPhit}
\begin{split}
\phi_s(\bk) & = \half \left( \phi_{\bk +} + \phi_{\bk -} \right) \\
\phi_t(\bk) & = \half \left( \phi_{\bk +} - \phi_{\bk -} \right) 
\end{split}
\eeq

This analysis sheds light on how an attraction in the singlet channel in presence
of a non-Abelian gauge field can produce a triplet piece in the pair
wave function. The triplet content $\etaT$ is now defined as the weight of the triplet piece of
the pair creation operator in \eqn{eqn:PairST}. One can also charecterize this by an expectation value of the quadrupole operator of reference \myonlinecite{Note2}. However, this definition for the triplet content $\etaT$ given above provides a physically transparent and a simple measure of the quantity of interest.

A remark about the Bogoliubov quasiparticles obtained in
\eqn{eqn:Bogoliubov} is in order. It appears that for each helicity
there are two branches of quasi-particle excitations labeled 1 and
2. Ostensibly, therefore, there are {\em four} branches of
quasiparticles which at the first sight is surprising. Note, however,
that these four branches are defined only in {\em half} of the
momentum space. If the Bogoliubov excitation were defined for all
$\bk$, they will not be independent, for example,
$\gamma^\dagger_{\bk2} \equiv \gamma^\dagger_{-\bk 1}$.  This is the
motivation behind introduction of the sum over one half of the
momentum space in \eqn{eqn:HalfSum}. It is now clear that the
formulation recovers the correct counting of excitation, i.e., within the
present formulation, two
excitations for each $\bk$ in momentum space is recovered as four
excitations for each $\bk$ in half the momentum space.

\begin{figure}
\centerline{{\large $~~~~~${\sf EP}, $\kf \as = -1$}}
\includegraphics[width=\myfigwidth]{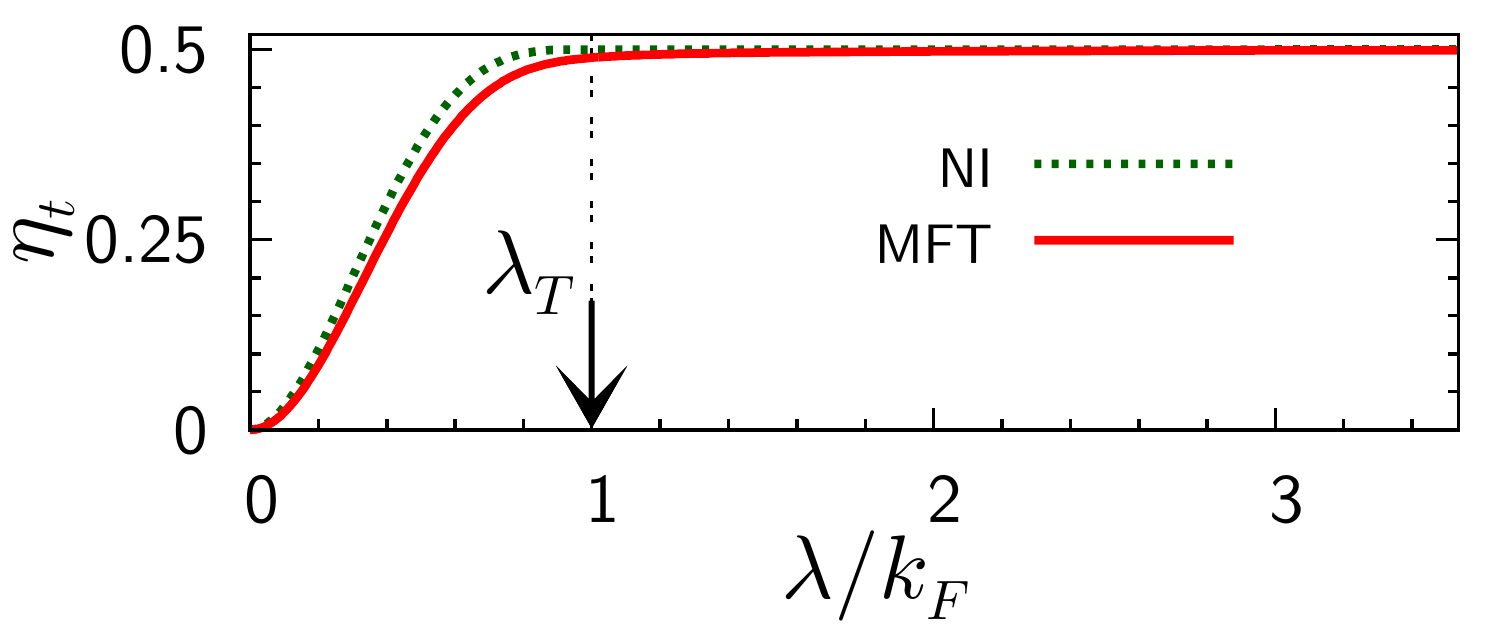}

\caption{(Color online) Evolution of the triplet content $\etaT$ of the pair wave
  function as a function of the gauge coupling strength $\lambda$ for
  an EP GFC with $\kf \as = -1$. The evolution of the same quantity of
  the non interacting system ($\as = 0^-$) is also shown for
  comparison.
}
\mylabel{fig:EPetaT}
\end{figure}

\begin{figure}
\centerline{{\large $~~~~~${\sf S}, $\kf \as = -\frac{1}{4}$}}
\includegraphics[width=\myfigwidth]{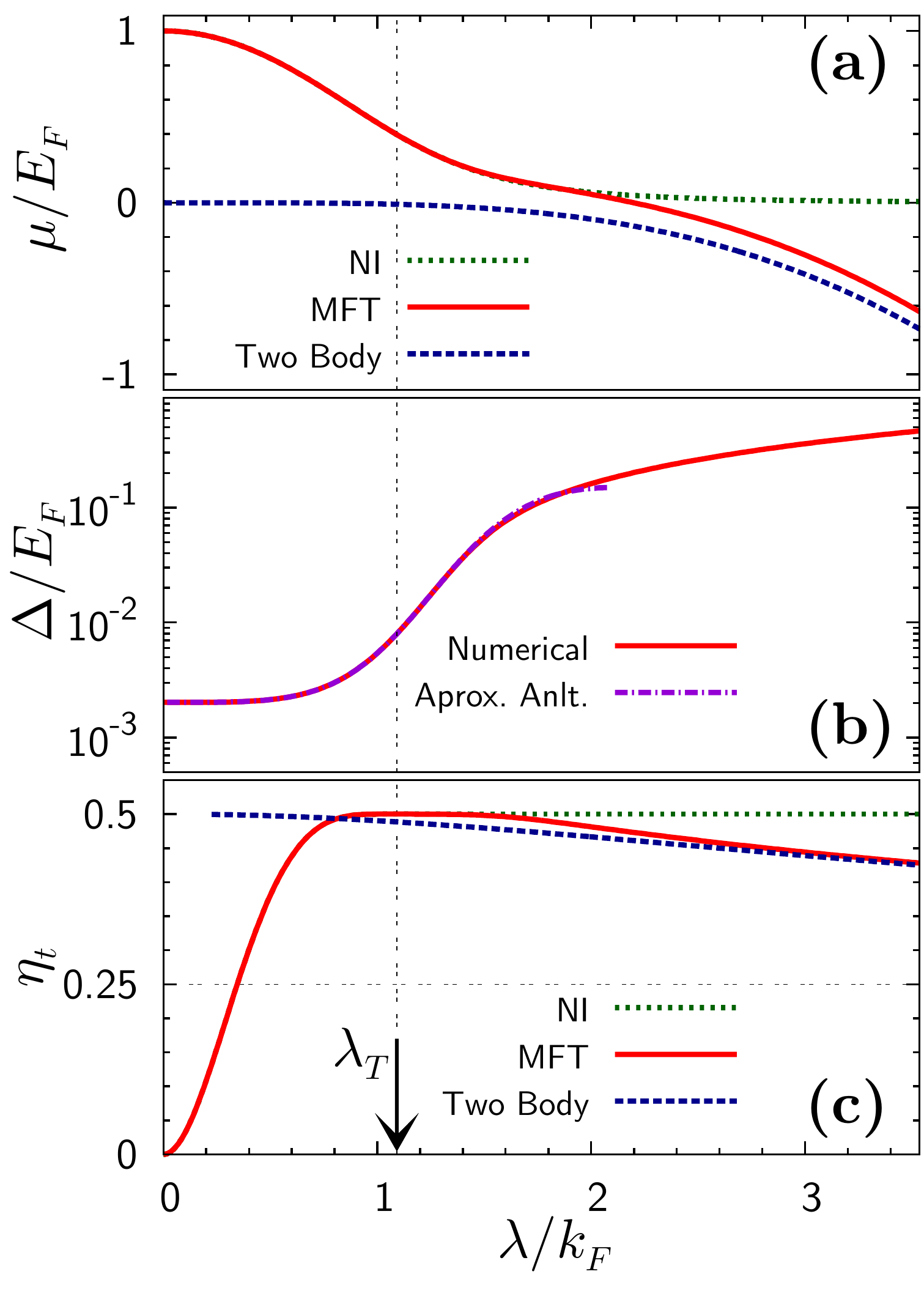}
\caption{(Color online) Evolution of the ground state of a collection of interacting
  fermions ($\kf \as = -\frac{1}{4}$) with gauge coupling strength
  $\lambda$ for the S GFC. {\bf (a)} Chemical potential obtained from a
  numerical solution of mean field theory (MFT) is compared with the
  chemical potential of the non interacting system (NI) and that set
  by the binding energy of the two body problem ($-\Eb/2$). For
  $\lambda \lesssim \lambda_T$ the chemical potential is
  indistinguishable from the that of the non-interacting system. For
  $\lambda \gtrsim \lambda_T$ the chemical potential approaches the
  two body value indicating a crossover to a BEC. {\bf (b)} Evolution
  of the numerically obtained mean field energy gap $\Delta$ with the
  gauge coupling strength $\lambda$. The analytical result
  \eqn{eqn:SAnltDelta} is also shown and is indistinguishable from the
  numerical result. {\bf (c)} The dependence of the triplet content ($\etaT$) of
  the pair wave function defined in \eqn{eqn:Pair} on the gauge
  coupling strength. This is compared with the same quantity of the
  non-interacting system (NI) and with that of the wave function of
  the two-body bound state. It is seen that the pair wave function
  evolves to two-body bound-state wave function.}
\mylabel{fig:SResults}
\end{figure}

\section{Results for specific Gauge Field Configurations}
\mylabel{sec:Results}

In this section, we shall present results of how the ground state of
the system evolves with $\lambda$ for various high symmetry GFCs. We
shall be concerned only with negative scattering lengths $(\as < 0)$
since this is the regime which has the most interesting physics. In
the absence of the gauge field $(\lambda =0)$ there is no two body
bound state, and for $\as < 0$ the usual BCS superfluid ground
state\cite{Randeria1995} $\ket{BCS_0}$ is obtained. For small
$\lambda$, i.~e., $\lambda \ll \lambda_T$, we expect and find the
ground state $\ket{\Psi_G}$ to be qualitatively close to $\ket{BCS_0}$ state
with an exponentially small excitation gap and a chemical potential
essentially unaltered from that of the non-interacting problem
$\muNI(\lambda)$. When $\lambda$ is increased beyond $\lambda_T$, we
find, in some cases (S and EO GFCs), that the chemical potential $\mu$
begins to fall and approaches $-\Eb/2$, the value set by energy of the
two body bound state. This signals the crossover to the BEC
state. Additionally, the pair wave function defined by \eqn{eqn:Pair}
approaches the wave function of the two body bound state.

A summary of the results for various GFCs discussed below is given in
Table.~\ref{tab:SummaryOfResults}.

\subsection{Extreme prolate (EP) GFC}
\mylabel{sec:EPResults}

This GFC with $\blam = (0,0,\lambda)$ has an FSTT at $\lambda_T =
\kf$. Before FSTT ($\lambda < \lambda_T$), the $+$ helicity Fermi sea
consists of the volume enclosed by two intersecting spheres of radius
$k_F$ centered around $(0,0,\pm\lambda)$, while the $-$ helicity Fermi
sea is the lens shaped region formed by the volume common to both
spheres. When $\lambda$ exceeds $\lambda_T$ the $-$ helicity Fermi
surface vanishes, and the $+$ helicity Fermi sea is made of two
disjoint spheres centered at $(0,0,\pm\lambda)$. The chemical
potential $\mu_{NI}(\lambda) = \Ef$, i.~e., is unaffected by the EP
gauge field.

\begin{figure}[h]
\centerline{{\large $~~~~~${\sf EO}, $\kf \as = -1$}}
\includegraphics[width=\myfigwidth]{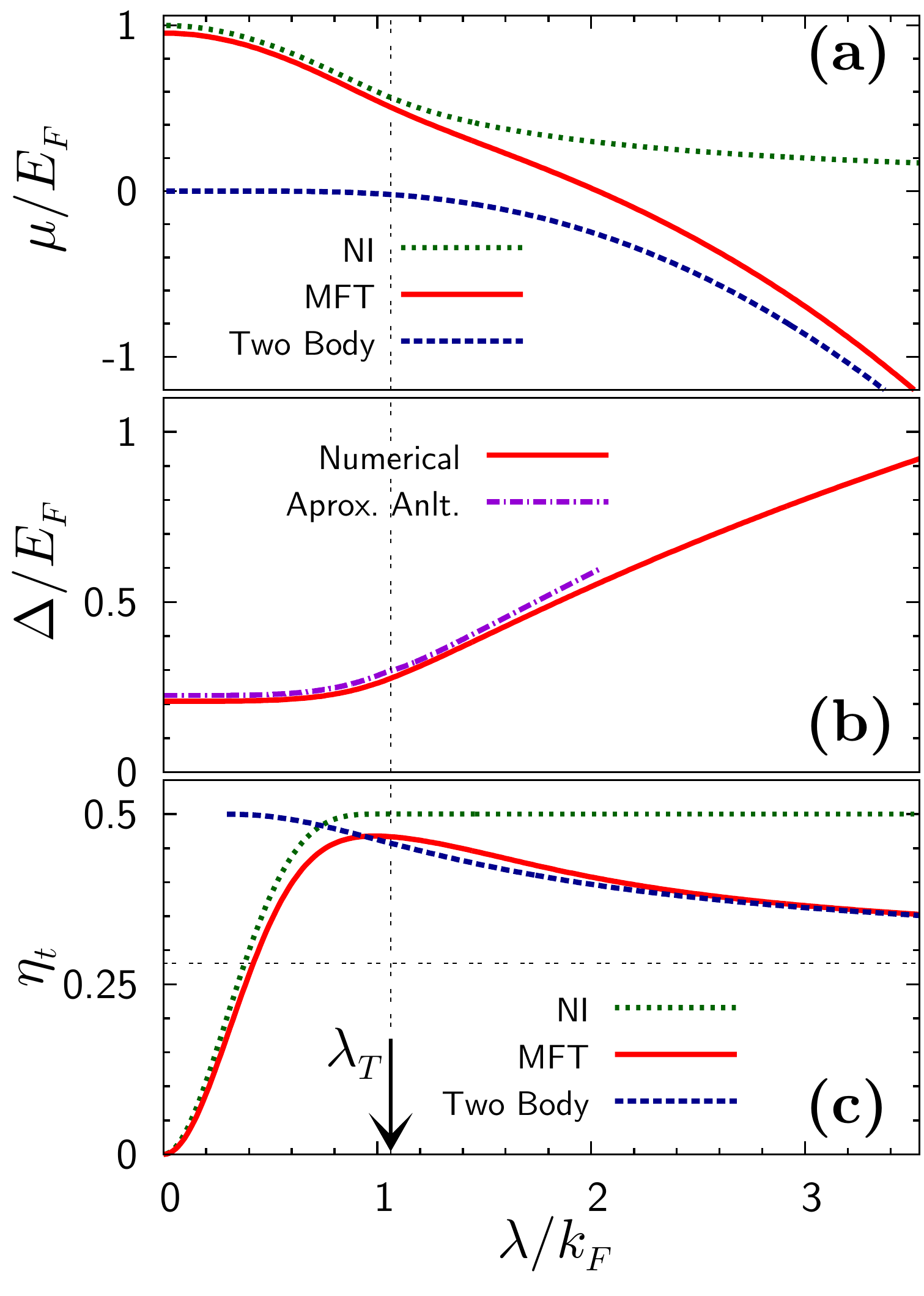}
\caption{(Color online) Evolution of the ground state of a collection of interacting
  fermions ($\kf \as = -1$) with gauge coupling strength $\lambda$ for
  the EO GFC. {\bf (a)} Chemical potential obtained from a numerical
  solution of mean field theory (MFT) is compared with the chemical
  potential of the non interacting system (NI) and that set by the
  binding energy of the two body problem ($-\Eb/2$). For $\lambda
  \lesssim \lambda_T$ the chemical potential has qualitative features
  similar to those the non-interacting system -- the numerical
  differences are due to the larger value of the magnitude of the
  scattering length. For $\lambda \gtrsim \lambda_T$ the chemical
  potential approaches the two body value indicating a crossover to a
  BEC. {\bf (b)} Evolution of the numerically obtained mean field
  energy gap $\Delta$ with the gauge coupling strength $\lambda$. The
  analytical result closely
  approximates the numerical result. {\bf (c)} The dependence of the
  triplet content ($\etaT$) of the pair wave function defined in \eqn{eqn:Pair}
  on the gauge coupling strength. This is compared with the same
  quantity of the non-interacting system (NI) and with that of the
  wave function of the two-body bound state. It is seen that the pair
  wave function evolves to two-body bound-state wave function.}
\mylabel{fig:EOResults}
\end{figure}

For $k_F |\as| \ll 1$, the standard result\cite{Randeria1995} for the excitation gap is
\beq
\frac{\Delta}{\Ef} \approx \frac{8 }{e^2} e^{-\frac{\pi}{2 k_F |\as|}} \mylabel{eqn:Gap0DeepBCS}
\eeq
and the chemical potential is
\beq
\mu \approx E_F \mylabel{eqn:Mu0DeepBCS}.
\eeq 
Not unexpectedly, the excitation gap $\Delta$ and the chemical
potential are {\em unaltered} with increasing $\lambda$. The ground
state for any $\lambda$ is a superfluid state with large overlapping
pairs, and {\em there is no BCS-BEC crossover} for the EP GFC. There is,
however, a qualitative change in the spin structure of the pair wave
function. With increasing $\lambda$, the pair wave function develops a
triplet content $\etaT$ (see \fig{fig:EPetaT}) which attains a value
close to $\half$ at $\lambda = \lambda_T$ and stays so with further
increase of $\lambda$.

The physics behind this result can be traced to the fact that for the
EP GFC the {\em kinetic energy content inside the non-interacting
  Fermi sea is unaltered by the increase of $\lambda$}. Therefore, the
gauge coupling $\lambda$ stays neutral in the competition between
kinetic energy and the attractive interaction. This, again, is the
reason why the energetics of the two-body problem is unaffected by the
presence of an EP gauge field (see
Table.~\ref{tab:TwoBodyInfo}). This is a feature specific only to EP GFCs.

It must be noted that the non-interacting ground state also has a triplet content (see
\Fig{fig:EPetaT}). As is evident (see \eqn{eqn:PairST}), this arises
from the fact that the $+$ helicity Fermi sea is different (and
larger) than the $-$ helicity Fermi sea. The triplet content of the
non-interacting system increases monotonically with $\lambda$ and
attains a value of $\half$ at $\lambda = \lambda_T$ and remains at
this value for any larger $\lambda$. As expected, in the presence of
an attractive interaction in the singlet channel ($\as < 0$), the
pairs have a triplet content {\em less} than that of the non
interacting system.

We note that the qualitative nature of the results for negative
scattering lengths ($\as < 0$) of larger magnitude are similar to
those for $k_F|\as| \ll 1$.

\subsection{Spherical (S) GFC}
\mylabel{sec:SResults}

When $\blam = \frac{\lambda}{\sqrt{3}}(1,1,1)$ a spherical (S) GFC is
obtained. Starting from two identical overlapping spheres at
$\lambda=0$, the non-interacting Fermi surfaces of the two helicities
continue to be spheres with their centers at the origin of the
momentum space for $0 < \lambda < \lambda_T$. Here $\lambda_T =
\displaystyle{\frac{\sqrt{3}}{2^{2/3}}\kf}$.  When $\lambda \ll
\lambda_T$, the chemical potential of the non-interacting system
depends on $\lambda$ as
\beq \mylabel{eqn:SmuNI}
\frac{\mu_{NI}(\lambda)}{E_F} = 1 - \frac{1}{2^{\frac{1}{3}}} \left(\frac{\lambda}{\lambda_T}\right)^2 \;\;\; (\lambda \ll \lambda_T)
\eeq
In this regime, the radius of the $+$ helicity Fermi surface is larger
than that of the $-$ helicity Fermi surface.  At the FSTT, the $-$
helicity Fermi surface vanishes and ceases to exist for all $\lambda
\ge \lambda_T$. After the FSTT, the $+$ helicity Fermi sea is ``a
sphere with a hole'', i.~e., the region bounded by two concentric spherical
Fermi surfaces. For $\lambda \gg \lambda_T$ the chemical potential of
the non-interacting system goes as
\beq
\frac{\mu_{NI}(\lambda)}{E_F} = \frac{2^{\frac{8}{3}}}{9} \left(\frac{\lambda_T}{\lambda}\right)^4 \;\;\; (\lambda \gg \lambda_T).
\eeq

Consider now the situation when $\kf |\as| \ll 1$. When $\lambda = 0$,
the usual BCS state with properties given by \eqn{eqn:Gap0DeepBCS} and
\eqn{eqn:Mu0DeepBCS} is the ground state. For $\lambda \ll \lambda_T$,
$\mu$ is very nearly equal to that given by \eqn{eqn:SmuNI}; the gap
equation can be solved analytically in this regime to obtain
\beq\mylabel{eqn:SAnltDelta}
\Delta = \frac{8 \mu_{NI}(\lambda)}{\exp{\left(\frac{12 \mu_{NI}(\lambda)}{6\mu_{NI}(\lambda)+\lambda^2}\right)}} \exp{\left(-\frac{3 \pi \sqrt{\mu_{NI}(\lambda)}}{\sqrt{2} |\as| (6\mu_{NI}(\lambda)+\lambda^2)}\right)}
\eeq

\Fig{fig:SResults}(a) and (b) show, respectively, the numerical
solutions of the chemical potential and gap as a function of
$\lambda$. \Fig{fig:SResults}(a) also shows the non-interacting
chemical potential, and the two-body energy $-\Eb/2$ (which depends on
$\lambda$ and $\as$ only). As is evident the chemical potential $\mu$
is identical to the non-interacting value $\muNI(\lambda)$ for
$\lambda \ll \lambda_T$.  There is also excellent agreement for the
gaps obtained from the numerical solution with the analytical result
given in \eqn{eqn:SAnltDelta}. When $\lambda$ reaches $\lambda_T$ the
chemical potential begins fall below $\mu_{NI}$, and {\em on further
  increase of $\lambda$ ($\lambda \gtrsim \lambda_T$), the chemical
  potential tends to that set by the two body problem.} This clearly
signals a crossover from the BCS like state for $\lambda \ll
\lambda_T$ to a BEC state where the fermions from tightly bound
bosonic pairs which then condense in the zero center of mass momentum
state.

Further corroboration of the crossover to the BEC like state with
increasing $\lambda$ can be obtained by a study of the triplet
fraction $\etaT$ which is shown in \Fig{fig:SResults}. Again, $\etaT$
corresponding to the non-interacting system monotonically increases
and attains a value of $\half$ at $\lambda_T$. The triplet content of
the superfluid pair, as expected, is less than that of the
non-interacting system, but has a similar qualitative behavior as the
NI case in the regime $\lambda \ll \lambda_T$. The triplet fraction
attains a maximum at a $\lambda$ close to $\lambda_T$ and then begins
to fall. On further increase of $\lambda$, $\etaT$ approaches that of
the {\em two-body} bound-state wave function, demonstrating again that
the pair wave function tends to the two-body bound-state wave
function. We also see that $\lambda= \lambda_T$ marks the crossover
point, i.~e., the crossover regime is precisely the regime of
$\lambda$ where change in the topology of the non-interacting Fermi
sea takes place.

It is particularly interesting to study the BEC state that is attained when
$\lambda \rightarrow \infty$.  The key point as noted in
section.~\ref{sec:ProbStatementAndSummary} is that the physics of the
two-body bound state is determined by the dimensionless parameter
$\lambda \as$ (see Table.~\ref{tab:TwoBodyInfo}). Therefore, as
$\lambda \rightarrow \infty$, the parameter $\frac{1}{\lambda \as}
\rightarrow 0$. Thus the state that is obtained is same as that
obtained for the two-body bound state with a resonant scattering
length {\em in the presence of the gauge field ($\lambda >0$)}
(Table.~\ref{tab:TwoBodyInfo})! Therefore {\em the properties of the
  BEC for $\lambda \rightarrow \infty$ are completely determined by
  $\blam$, independent of the scattering length (as long as it is non
  vanishing),} i.~e., the system is a collection of Bosons whose
properties are determined solely by the Rashba interaction. Hence we
call this tightly bound bosonic state of two fermions as ``rashbon''. Rashbon is a bound state of two fermions in a Rashba gauge field ($\lambda>0$) at resonant scattering length ($\frac{1}{\as}=0$).

Again, for scattering lengths of larger magnitude, the qualitative
physics remains identical. We shall illustrate this point in the next
section by considering the EO case with a scattering length of larger
magnitude.

\subsection{Extreme oblate (EO) GFC}
\mylabel{sec:EOResults}

The evolution of the non interacting Fermi surfaces for this GFC
$(\blam = \frac{\lambda}{\sqrt{2}} (1, 1, 0)$ is shown in
\fig{fig:EOTopologicalTransition}. The non-interacting chemical
potential in the regime $\lambda \ll \lambda_T$ is
\beq\mylabel{eqn:EOmuNI}
\frac{\mu_{NI}(\lambda)}{E_F} = 1 - \left(\frac{4}{3\pi}\right)^{\frac{2}{3}} \left(\frac{\lambda}{\lambda_T}\right)^2 \;\;\; (\lambda \ll \lambda_T)
\eeq
and that in the regime  $\lambda \gg \lambda_T$ is
\beq
\frac{\mu_{NI}(\lambda)}{E_F} = \left(\frac{4}{3\pi}\right)^{\frac{2}{3}} \frac{\lambda_T}{\lambda} \;\;\; (\lambda \gg \lambda_T).
\eeq

In the regime when $\lambda \ll \lambda_T$ and for $\kf |\as| \ll 1$,
the chemical potential is well approximated by the non-interacting
value. Further, we can obtain a rather lengthy analytical expression
for the gap (not shown).

To illustrate that the qualitative nature of the transition is
unaltered by the size of the scattering length, we study this GFC with
$\kf|\as| =1$. The results are shown in \fig{fig:EOResults}. These
results clearly illustrate a crossover from the BCS like state to a
BEC state of rashbons. Note, in particular, that $\etaT$ of the many
body pair wave function tends to that of the two body bound state wave
function with a resonant scattering length (rashbon) given in
Table.~\ref{tab:TwoBodyInfo}.

\section{Discussion}
\mylabel{sec:Discussion}

We conclude the paper with further discussion of our results. On the
BCS side $\kf|\as| \ll 1$ and $\lambda \ll \lambda_T$, the transition
temperature will be determined by the zero temperature gap which we
have calculated in this paper. On the rashbon BEC side, the transition
temperature will be determined by the mass of these emergent
bosons\cite{Vyasanakere2011a} which will be renormalized from the
value of twice the fermion mass due to the gauge field. 

As noted earlier, the rashbon is a bound state of two fermions
  in a Rashba gauge field $(\lambda >0)$ when the $s$-wave scattering
  length is infinity, i.e., at resonance. This two-fermion bound state
  exists for {\em all} GFCs except the EP GFC and has a spin
structure determined by $\hat{\blam}$ of the GFC (see the ``resonance'' column of
Table.~\ref{tab:TwoBodyInfo}). As is evident this state is not rotationally symmetric -- it is an ``anisotropic particle'' that emerges. It is also interesting
to contrast the rashbon state obtained in a Rashba gauge field
($\lambda > 0)$ with the two-body quasi-bound state obtained in free
vacuum ($\lambda=0)$ at resonance. In the latter case, the binding energy is zero, and the state
is scale free with a singlet spin structure. This is to be contrasted
with the rashbon state whose binding energy is $\lambda^2$ times a
dimensionless number that depends on $\hat{\blam}$. Indeed, the state is not scale free -- the wave function in the relative coordinate of the two
fermions dies exponentially with a scale $\lambda^{-1}$ as noted in
reference \myonlinecite{Vyasanakere2011}.

For a generic GFC, it is known that the critical scattering length
$\asc$ required to induce a bound state is negative and
finite\cite{Vyasanakere2011} and is given by $\asc =
\frac{{\cal{F}}(\hat{\blam})}{\lambda}$ where ${\cal F}$ is a
dimensionless function. For a {\em given} $\as < 0$, this corresponds
to a critical gauge coupling strength $\displaystyle{\lambda_c =
  \left|\frac{{\cal{F}}(\hat{\blam})}{\as}\right|}$. The crossover
with increasing $\lambda$ is then  governed by the relative
magnitudes of $\lambda_T$ and $\lambda_c$. If $\lambda_c \lesssim
\lambda_T$, the crossover regime  coincides with the regime of the
FSTT. On the other hand if $\lambda_c \gg \lambda_T$, the crossover
regime is centered  around $\lambda \approx \lambda_c$. In any case, for
$\lambda \gg \max{(\lambda_T,\lambda_c)}$ the ground state will be a
condensate of rashbons determined by the GFC
in question. It is evident that except for the EP GFC, every other GFC
will support a BCS-rashbon BEC crossover.

We now discuss the situation with a small positive
scattering length with $\kf \as \ll 1$. In absence of a gauge field, the
ground state is BEC of bosonic pairs of fermions with mass twice that
of the fermion mass. In the presence of the gauge field, this BEC will
evolve to the rashbon BEC as $\lambda \rightarrow \infty$, i.e, there
is a BEC-rashbon BEC crossover.

The authors are not aware of any experimental realization of synthetic
gauge field in fermionic systems. The natural
question that arises is if the parameter regime of $\lambda \gtrsim
\lambda_T$ with a high symmetry GFC can be realized in experiments. We do hope that our paper provides the motivation for this
direction of experimental research.

\subsection*{Acknowledgement}

JV acknowledges support from CSIR, India via a JRF grant. SZ is supported by NSF DMR-0907366 and DARPA Nos. W911NF-07-1-0464
 . VBS is grateful to DST, India (Ramanujan grant) and DAE,
India (SRC grant) for generous support. We are grateful to Tin-Lun
(Jason) Ho for discussions.

\bibliography{refbcsbec}

\end{document}